\documentclass[aip,rsi,amsmath,amssymb,reprint]{revtex4-1}

\usepackage{graphicx}
\usepackage{dcolumn}
\usepackage{bm}

\usepackage[utf8]{inputenc}
\usepackage[T1]{fontenc}
\usepackage{mathptmx}
\usepackage{etoolbox}
\usepackage{siunitx}
\usepackage{old-arrows}

\makeatletter
\def\@email#1#2{%
 \endgroup
 \patchcmd{\titleblock@produce}
  {\frontmatter@RRAPformat}
  {\frontmatter@RRAPformat{\produce@RRAP{*#1\href{mailto:#2}{#2}}}\frontmatter@RRAPformat}
  {}{}
}
\makeatother

\DeclareSIUnit\ion{ions}
\DeclareSIUnit\gauss{g}

\begin{document}

\title{Homogeneous Microwave Delivery for Quantum Sensing with Nitrogen-Vacancy Centers at High Pressures}
\author{Timothy A. Elmslie}
\affiliation{Sandia National Laboratories, Albuquerque, New Mexico 87185, USA}
\author{Luca Basso}
\affiliation{Center for Integrated Nanotechnologies, Sandia National Laboratories, Albuquerque, New Mexico 87123, USA}
\author{Adam Dodson}
\affiliation{Center for Integrated Nanotechnologies, Sandia National Laboratories, Albuquerque, New Mexico 87123, USA}
\author{Jacob Henshaw$^*$}
\email{jhensha@sandia.gov}
\affiliation{Sandia National Laboratories, Albuquerque, New Mexico 87185, USA}
\author{Andrew M. Mounce$^*$}
\email{amounce@sandia.gov}
\affiliation{Center for Integrated Nanotechnologies, Sandia National Laboratories, Albuquerque, New Mexico 87123, USA}

\date{\today}
\preprint{AIP/123-QED}

\begin{abstract}
Nitrogen vacancy (NV) centers have been demonstrated as a useful tool in high pressure environments. However, the geometry and small working area of the diamond anvil cells (DACs) used to apply pressure present a challenge to effective delivery of microwave (mw) fields. We designed and characterized a novel slotted design for mw transmission to nitrogen-vacancy centers (NVs) in a diamond anvil cell via zero-field and in-field optically detected magnetic resonance (ODMR) measurements across pressures between 1 and~\SI{48}{\giga \pascal}.  The mw fields experienced by NVs across the diamond culet was calculated from Rabi frequency and found to be higher and more uniform than those generated by an equivalent simple mw line, which will improve performance for wide-field, high-pressure measurements to probe spatial variations across samples under pressure.
\end{abstract}
\maketitle

\section{Introduction}
Diamond anvil cells (DACs) have been used extensively for high-pressure experiments since their invention by Weir~\textit{et al.}~\cite{weir_infrared_1959} in the late 1950s~\cite{jayaraman_diamond_1983,hamlin_superconductivity_2007,bassett_diamond_2009,qin_review_2024,struzhkin_magnetic_2024}.
Diamonds' hardness allows them to withstand high pressures without deforming themselves, and they are transparent to many wavelengths of light, permitting optical measurements on a pressurized sample.
Combined with techniques for four-wire resistivity and AC susceptibility measurements under pressure, DACs have been instrumental in the search for high-temperature superconductors, finding superconducting transitions at temperatures as high as~\SI{203}{\kelvin} in H$_3$S~\cite{drozdov_conventional_2015} compressed to~\SI{155}{\giga \pascal} and approximately~\SI{250}{\kelvin} in LaH$_{10-\delta}$ at~\SI{200}{\giga \pascal}~\cite{somayazulu_evidence_2019,drozdov_superconductivity_2019}.

Negatively-charged nitrogen-vacancy centers (NVs) in diamond have been the subject of numerous studies due to their usefulness in quantum sensing~\cite{maze_nanoscale_2008,doherty_nitrogen-vacancy_2013,rondin_magnetometry_2014,abe_tutorial_2018,levine_principles_2019,barry_sensitivity_2020,xu_recent_2023}.
The optical cycle of the NV center allows for optical readout of the electronic spin state, creating a non-invasive probe of its magnetic environment.
Naturally, NV magnetometry has found use in DACs for sensing under pressure, in which one diamond anvil can be implanted with NVs, or NV-implanted diamond powder can be included in the cell alongside the sample~\cite{doherty_electronic_2014,steele_optically_2017,shang_magnetic_2019,hsieh_imaging_2019,yip_measuring_2019,lesik_magnetic_2019,ho_probing_2020,dai_optically_2022,ho_spectroscopic_2023,hilberer_enabling_2023,bhattacharyya_imaging_2024}.
This technique has proven particularly useful in the detection of bulk superconductivity under pressure via the Meissner effect~\cite{bhattacharyya_imaging_2024,dailledouze_imaging_2025}, as prior methods such as four-wire resistivity measurements require extremely precise placement of electrical leads on the sample, while AC susceptibility measurements via pickup coil lack spatial resolution and can potentially detect signals other than those of the measured sample~\cite{ho_recent_2021}.

The small working area of a DAC presents experimental challenges, however, and a number of solutions have emerged in response to the problem of microwave delivery, such as loops or coils of wire around or on the culet~\cite{doherty_electronic_2014,ho_spectroscopic_2023}, coils embedded within epitaxial diamond underneath the culet~\cite{steele_optically_2017}, using a loop of wire on a split gasket~\cite{lesik_magnetic_2019,hilberer_enabling_2023}, or a coil inside the sample chamber~\cite{ho_probing_2020,yip_measuring_2019}.
A particularly common method is a strip of Pt foil or thin Pt wire placed on the gasket near the sample~\cite{hsieh_imaging_2019,bhattacharyya_imaging_2024,mai_megabar_2025,he_probing_2025}, since this creates a magnetic field parallel to the culet surface and allows the use of [111]-oriented diamond, but this method results in an inhomogeneous mw field which decreases as $\frac{1}{r}$ across the culet.
The magnitude of the mw field is directly related to Optically Detected Magnetic Resonance (ODMR) contrast and therefore sensitivity, so a strong and uniform field across the sample is desirable.
This paper demonstrates that such a field can be achieved simply by cutting a rectangular slot into a strip of Pt foil and placing this slot over the gasket hole of a DAC.

\section{Methods}
\begin{figure}
        \centering
        \includegraphics[width=0.8\linewidth]{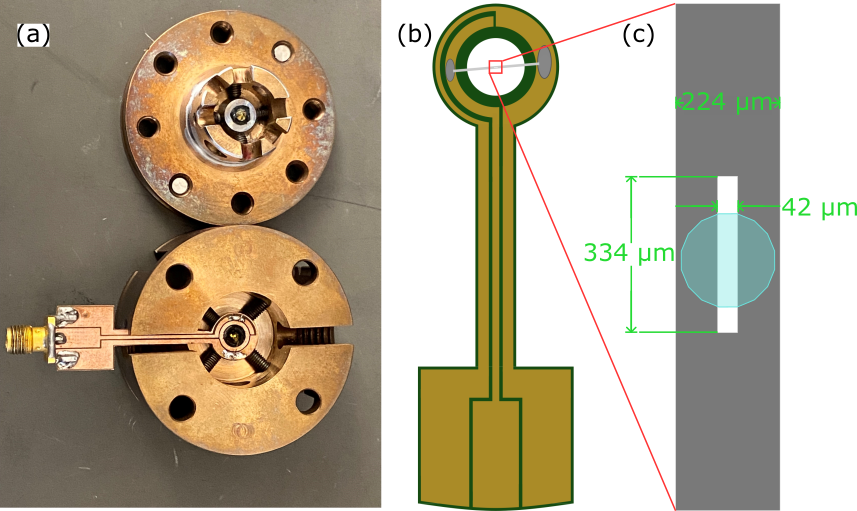}
        \caption{(a) The modified Diacell SymmDAC 60 BeCu showing the vertical slot, PCB, and slotted mw line.  (b) Detailed diagram of the PCB used to connect the mw line to an external mw source. (c)  Detailed diagram of the slotted mw line.}
        \label{fig:1}
    \end{figure}

    Measurements were performed in a Diacell SymmDAC 60 BeCu from Almax modified with a vertical slot in both the piston and cylinder sides of the cell as shown in Fig.~\ref{fig:1}(a).
    These vertical slots allow the use of a PCB to connect a mw line in the cell to an external mw source to drive the $m_s = 0 \leftrightarrow -1$ transition of the NV center.
    Figure~\ref{fig:1}(b) shows the design of the PCB, and (c) contains a diagram of the mw line.
    
    We placed two type Ib diamonds cut in the [111] direction with \SI{200}{\micro \meter} culets in the cell to apply pressure, each containing roughly 200 ppm N.
    One of these diamonds was implanted via bombardment with~\SI{10}{\kilo \electronvolt} $^{12}$C atoms at a fluence of~\SI{1e13}{\ion \per \centi \meter \squared} resulting in an implant depth of about~\SI{10}{\nano \meter} with a vacancy concentration of~\SI{200} ppm. The implanted diamond is then annealed to form NVs and cleaned to oxygen terminate the diamond surface following the methods laid out in previous work by J. Henshaw \textit{et al.}~\cite{henshaw_nanoscale_2022}.
    The NVs were excited with a 532 nm laser and fluorescence was measured using a microscope and optical setup also similar to J. Henshaw \textit{et al.}~\cite{henshaw_nanoscale_2022}. , though the microscope objective was replaced with an NA = 0.4, 20$\times$~objective providing a beam diameter of 5 $\mu$m.
    Experiments were controlled via the QICK-DAWG open source quantum hardware framework~\cite{riendeau2023quantuminstrumentationcontrolkit}. 
    A rhenium gasket was preindented to a thickness of roughly~\SI{25}{\micro \meter} and insulated using alumina powder.
    A laser drill was used to bore a sample space with a roughly~\SI{60}{\micro \meter} diameter in the center of the preindented area.
    This sample space was filled with Si oil as a pressure transmitting medium. 
    The slotted platinum foil for mw delivery was prepared using a laser drill.
    Before applying pressure, the foil was about~\SI{4}{\micro \meter} thick, and~\SI{224}{\micro \meter} wide, with a slot that was~\SI{42}{\micro \meter} wide.
    Measurements were performed without a sample to characterize the mw field experienced by the NVs.
    
\section{Results \& Discussion}    
\begin{figure}
        \centering
        \includegraphics[width=1.0\linewidth]{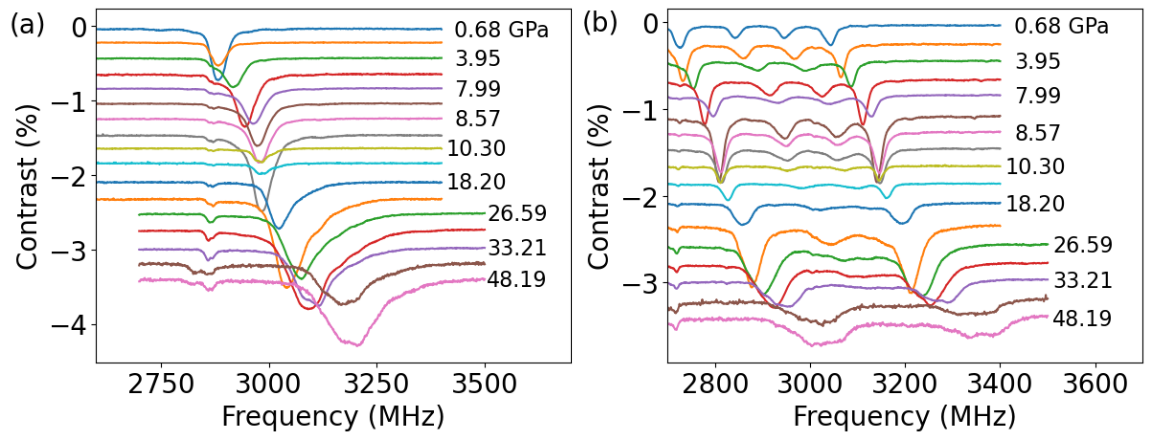}
        \caption{Optically-detected magnetic resonance (ODMR) spectra at various pressures. Plot (a) shows results without an applied static magnetic field, and plot (b) shows results for an applied field of approximately~\SI{5.6}{\milli \tesla}.  In both plots, datasets are vertically offset for clarity.}
        \label{fig:ODMRWaterfall}
    \end{figure}

    Optically-detected magnetic resonance (ODMR) spectra are shown in Fig.~\ref{fig:ODMRWaterfall} across a range of pressures.
    In both plots, spectra are displaced on the y-axis for clarity, with lower position corresponding to increasing pressure.
    Pressures were determined via ruby fluorescence using the calibration generated by Dewaele~\textit{et al.}~\cite{dewaele_compression_2008}.
    In all cases, measurements were performed within~\SI{2}{\micro \meter} of the center of the mw slot. Figure~\ref{fig:ODMRWaterfall}(a) contains spectra obtained without an applied static magnetic field, while for Fig.~\ref{fig:ODMRWaterfall}(b), a small permanent magnet was placed under the cell to provide a field $B_0$ perpendicular to the diamond culets.
    This field splits the ODMR spectrum into four peaks.
    This behavior can be explained by the Hamiltonian of the NV center in its ground state
    \begin{equation}
        H = \vec{S} \cdot \overleftrightarrow{D} \cdot \vec{S} + \gamma_e \vec{B}_0 \cdot \vec{S},
    \end{equation}
    in which $\vec{S}$ is the electron spin operator, $\overleftrightarrow{D}$ is the spin-spin coupling tensor, $\gamma_e$ is the electron gyromagnetic ratio of about~\SI{28}{\giga \hertz \per \tesla}, and $\vec{B}_0$ is the external magnetic field~\cite{ho_recent_2021}.
    Due to the [111] cut of the diamonds, the field projection $\vec{B}_0 \cdot \vec{S}$ onto the [111] NV axis is maximized, while the field projection onto each of the other NV axes is smaller and roughly equivalent.
    Therefore, for the [111] axis, we approximate $\overleftrightarrow{D}$ as only the $S_z^2$ term, making it roughly equal to~\SI{2.87}{\giga \hertz} at ambient conditions.
    Since the Zeeman splitting is related to this projection, four peaks appear on the spectrum, with the maximally-displaced outer two peaks corresponding to the [111]-oriented NVs and the inner two originating from the other axes~\cite{mai_megabar_2025}.
    The field experienced by the NVs can be estimated from the splitting of the [111] peaks according to~\cite{abe_tutorial_2018,ho_recent_2021}
    \begin{equation}
        B_0 = \frac{\Delta f}{2 \gamma_e},
    \end{equation}
    in which $\Delta f$ is frequency difference between the $0 \leftrightarrow -1$ and $0 \leftrightarrow +1$ transition peaks and $2\gamma_e$ is approximately~\SI{5.6}{\mega \hertz \per \gauss}.
    In this case, the calculation produces a $B_0$ field of about~\SI{5.6}{\milli \tesla}.
    The ODMR peaks shift to higher frequency as pressure increases due to the compression of the electron wave functions within the NV center~\cite{doherty_electronic_2014} resulting in an increase of the zero-field splitting.
    Both [111] peaks are affected equally, and the frequency shift can be quantified by simply averaging the peak positions.
    \begin{figure}
        \centering
        \includegraphics[width=1.0\linewidth]{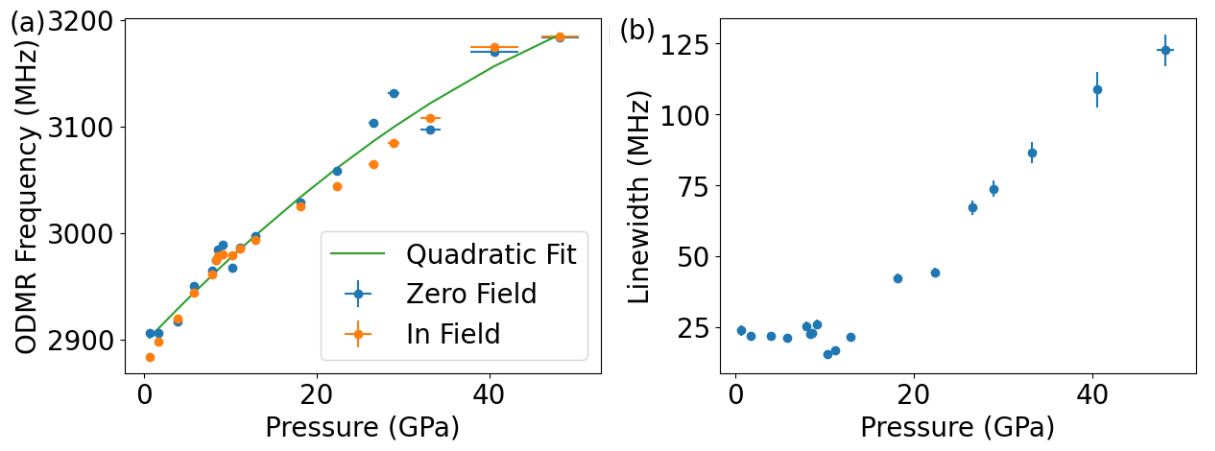}
        \caption{(a) ODMR frequency versus pressure values measured with applied static magnetic field (yellow points) and without static field (blue points).  A quadratic fit $\nu = a + bP + cP^2$ to the data is shown in green, in which $a$ is approximately~\SI{2896(5)}{\mega \hertz}, $b$ is~\SI{8.5(6)}{\mega \hertz \per \giga \pascal}, and and $c$ is about~\SI{-0.052(1)}{\mega \hertz \per \giga \pascal \squared}.  (b) Linewidth versus pressure values calculated via Lorentzian fits to in-field ODMR peaks.}
        \label{fig:FreqvsP}
    \end{figure}
    Plotting this value as a function of pressure gives the plot in Fig.~\ref{fig:FreqvsP}(a); zero field data are plotted in blue, with in-field in orange.
    Fitting the equation $\nu = a + bP + cP^2$ to the combined in-field and zero-field data produces the parameters $a=\SI{2896(5)}{\mega \hertz}$, $b=\SI{8.5(6)}{\mega \hertz \per \giga \pascal}$, and $c=\SI{-0.052(1)}{\mega \hertz \per \giga \pascal \squared}$.
    This fit is shown as the green curve in Fig.~\ref{fig:FreqvsP}(a).
    These results are comparable to those obtained by D.P. Shelton~\textit{et al.}~\cite{shelton_magnetometry_2024}, who observed $a=\SI{2871(1)}{\mega \hertz}$, $b=\SI{15.5(2)}{\mega \hertz \per \giga \pascal}$, and $c=\SI{-0.044(7)}{\mega \hertz \per \giga \pascal \squared}$ in randomly-oriented nanodiamond particles in a DAC.

    We performed a Lorentzian fit on in-field [111] ODMR peaks to obtain linewidth at each pressure.
    The resulting plot is displayed in Fig.~\ref{fig:FreqvsP}(b).
    Linewidth remains mostly constant until about~\SI{10}{\giga \pascal}, at which point it slightly decreases.
    At roughly~\SI{12}{\giga \pascal}, linewidth begins increasing linearly up to the highest measured pressures.
    This plot bears notable similarity to the plot of standard deviation in pressure measured in Si oil obtained by Klotz~\textit{et al.}~\cite{klotz_hydrostatic_2009}, including the decrease near~\SI{10}{\giga \pascal}, suggesting that increases in linewidth are largely due to pressure gradients caused by the Si oil pressure medium.

    The value of $T_2^*$ was calculated according to the method of Acosta~\textit{et al.}~\cite{acosta_diamonds_2009} by measuring linewidth as a function of microwave power and extrapolating to zero power, although we employed a power law function rather than the linear fit used by Acosta~\textit{et al.}
    Details are presented in the supplementary material.
    Our $T_{2}^{*}$ was found to be~\SI{19.1(1)}{\nano \second}, similar in magnitude to the expected value of about~\SI{50}{\nano \second} for a diamond containing 200 ppm N~\cite{bauch_decoherence_2020}.

    \begin{figure}
        \centering
        \includegraphics[width=1.0\linewidth]{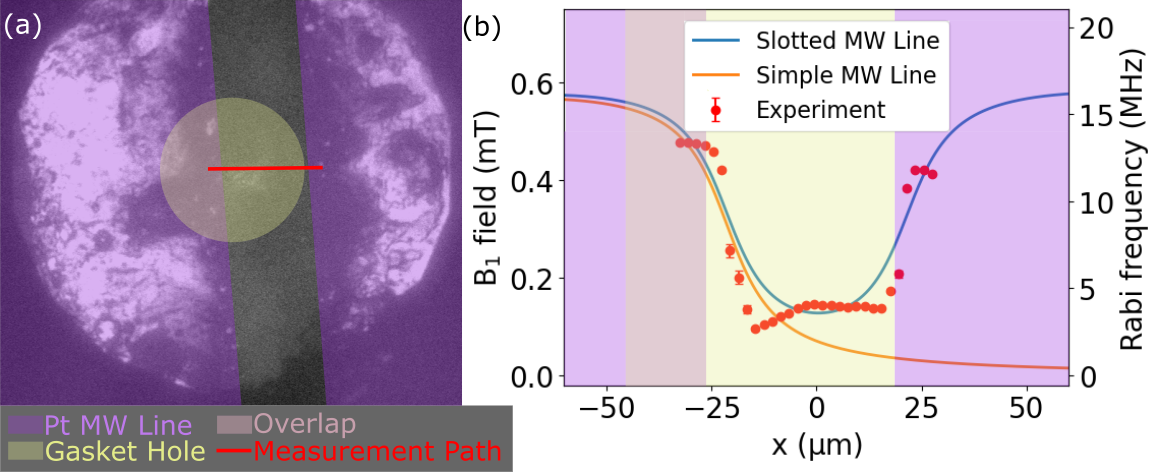}
        \caption{(a) Picture of diamond culet with Pt mw line highlighted in purple and the gasket hole highlighted in yellow.  A red line shows the approximate path of the Rabi measurements that were taken across the slot of the mw line.  (b) Microwave field strength measured across the mw line at ambient pressure.  Experimental values calculated from Rabi frequency measurements are shown in red, with theoretical results for slotted and non-slotted mw lines in blue and orange respectively.  Purple and yellow highlights show the approximate positions of the mw line and gasket hole relative to the measurements taken.}
        \label{fig:SimVSExp}
    \end{figure}
    Figure~\ref{fig:SimVSExp}(a) displays a picture of the diamond culet and mw line, with the line highlighted in purple for clarity.
    The gasket hole containing the pressure medium is difficult to see due to the brightness of the Pt mw line in the image, but its approximate location is highlighted in yellow.
    Rabi frequency measurements were taken across the mw line in roughly~\SI{2}{\micro \meter} steps; this path is shown in red in Fig.~\ref{fig:SimVSExp}(a).
    and used to calculate the mw field experienced by the NVs at each position according to the equation $f=\frac{f_{Rabi}}{\gamma_e}$, in which $f_{Rabi}$ is the Rabi frequency and $\gamma_e$ is the electron gyromagnetic ratio. 
    The results of this measurement are plotted in Fig.~\ref{fig:SimVSExp}(b) as red points.
    A simulation of the mw field experienced by the [111]-oriented NVs from the slotted mw line is plotted in blue.
    The orange curve represents a simulation of the field produced by a non-slotted ordinary mw line placed beside the sample.

    The slotted microwave wire provides a much more uniform driving field for the NVs.
    The measured magnetic field $B_1$ is nominally flat over the sample volume and avoids the 1/r drop off anticipated from a single wire shown in Fig~\ref{fig:SimVSExp}(b). 
    Additionally, the measured Rabi frequencies are on the order of 3 MHz, offering enough bandwidth to cover the spectrum for an NV Center with $^{15}$N (Spin-1/2, hyperfine of 3.05 MHz), but just missing the bandwidth requirements for an NV with $^{14}$N (Spin-1 hyperfine of 2.2 MHz).
    With more optimization, this mw field strength can be increased. 
    Since the slot in the wire needs to be slightly larger than the sample volume, no additional special tools are needed for making the slot.
    
    The slotted wire design does impose some limitations on conventional DAC measurements. 
    Since the slot encompasses the sample volume, performing resistance measurements on a sample in the sample volume is made more challenging because any leads will short to the microwave line without proper precaution.
    
\section{Conclusion}
We have demonstrated the design and use of a novel method of mw delivery for NV magnetometry in a DAC and shown its performance through ODMR measurements and $B_1$ field calculated from Rabi frequency.
We also obtained linewidth as a function of pressure, demonstrating that sensitivity remains nearly constant until about~\SI{10}{\giga \pascal}, after which it worsens likely due to increasing anisotropic stresses.
While Si oil pressure medium becomes non-hydrostatic at roughly~\SI{0.9}{\giga \pascal}~\cite{angel_effective_2007}, pressure anisotropy remains relatively low until about~\SI{12}{\giga \pascal}, at which point the oil undergoes a phase transition into a glass~\cite{chervin_diamond_1995,klotz_hydrostatic_2009}.
By comparing experimentally-determined mw field with theoretical calculations, we showed that the slotted mw line creates a more uniform field across the sample space than a simple strip of foil.
Therefore, the slotted mw line may prove beneficial to spatially-resolved NV measurements in particular, which have recently been used to examine superconducting transitions~\cite{bhattacharyya_imaging_2024,dailledouze_imaging_2025}.

In future work, performance of NVs driven by the slotted mw line could be further improved by experimenting with different nitrogen concentrations and implant parameters as well as the mw line dimensions.
Additionally, embedding a tungsten mw line within epitaxial diamond may solve any difficulties the present design might create for transport measurements.
Such a designer diamond anvil might also allow a wider range of mw line designs.
For instance, numerous, equally-spaced thin strips of tungsten across the culet could create an even more uniform field than the slotted mw line that we present while still allowing optical access to the sample.

\section{Supplementary Material}
See the supplementary material for detailed procedures on the assembly of the diamond anvil cell and mw line used for the measurements presented here.

\section{Acknowledgments}
The authors thank Sakun Duwal, Vinnie Martinez, Ryan Schecker, and Benjamin Halls for the use of scientific equipment and facilities. The authors also thank Brian Maple, Eric Lee Wong, and Yuhang Deng for providing platinum foil and valuable discussion.
Sandia National Laboratories is a multimission laboratory managed and operated by National Technology \& Engineering Solutions of Sandia, LLC, a wholly owned subsidiary of Honeywell International Inc., for the U.S. Department of Energy’s National Nuclear Security Administration under contract DE-NA0003525. SAND2026-15944O.
This written work is authored by an employee of NTESS. The employee, not NTESS, owns the right, title and interest in and to the written work and is responsible for its contents.
Any subjective views or opinions that might be expressed in the written work do not necessarily represent the views of the U.S. Government. The publisher acknowledges that the U.S. Government retains a non-exclusive, paid-up, irrevocable, world-wide license to publish or reproduce the published form of this written work or allow others to do so, for U.S. Government purposes. The DOE will provide public access to results of federally sponsored research in accordance with the DOE Public Access Plan.
This work was funded, in part, by the Laboratory Directed Research and Development Program and performed, in part, at the Center for Integrated Nanotechnologies, an Office of Science User Facility operated for the U.S. Department of Energy (DOE) Office of Science.

\section{Author Declarations}
\subsection{Conflict of Interest}
The authors have no conflicts of interest to disclose.

\subsection{Author Contributions}
\textbf{Timothy A. Elmslie:} Investigation, Methodology, Writing - original draft, Writing - review \& editing.
\textbf{Luca Basso:} Methodology.
\textbf{Adam Dodson:} Methodology.
\textbf{Jacob Henshaw:} Methodology, Writing - review \& editing, Supervision.
\textbf{Andrew M. Mounce:} Conceptualization, Methodology, Supervision.

\section{Data Availability}
The data that support the findings of this study are available from the corresponding author upon request.
\section{References}
\bibliographystyle{unsrt}
\bibliography{zotero_references,references}
\end{document}


\renewcommand{\theequation}{S\arabic{equation}}
\renewcommand{\thefigure}{S\arabic{figure}}
\renewcommand{\bibnumfmt}[1]{[S#1]}
\renewcommand{\citenumfont}[1]{S#1}
\title{Supplementary Material: Homogeneous Microwave Delivery for Quantum Sensing with Nitrogen-Vacancy Centers at High Pressures}

\author{Timothy A. Elmslie}
\affiliation{Sandia National Laboratories, Albuquerque, New Mexico 87185, USA}
\author{Luca Basso}
\affiliation{Center for Integrated Nanotechnologies, Sandia National Laboratories, Albuquerque, New Mexico 87123, USA}
\author{Adam Dodson}
\affiliation{Center for Integrated Nanotechnologies, Sandia National Laboratories, Albuquerque, New Mexico 87123, USA}
\author{Jacob Henshaw$^*$}
\email{jhensha@sandia.gov}
\affiliation{Sandia National Laboratories, Albuquerque, New Mexico 87185, USA}
\author{Andrew M. Mounce$^*$}
\email{amounce@sandia.gov}
\affiliation{Center for Integrated Nanotechnologies, Sandia National Laboratories, Albuquerque, New Mexico 87123, USA}

\date{\today}
\preprint{AIP/123-QED}
\maketitle

\section{Rabi Oscillation}
\begin{figure}
    \centering
    \includegraphics[width=0.9\linewidth]{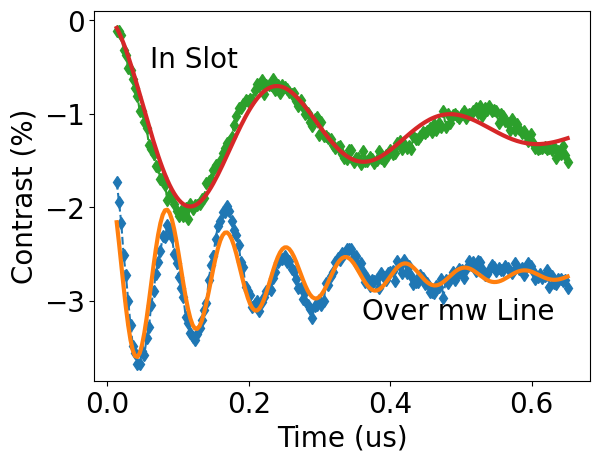}
    \caption{Examples of Rabi measurements taken across the mw line.  Green datapoints were taken near the center of the mw line, while blue were taken on the Pt material of the mw line.  Red and orange lines show the fits to each of the respective datasets.  Datasets are offset for clarity.}
    \label{fig:RabiFit}
\end{figure}

Rabi frequency measurements were used to determine mw field strength at various positions across the mw line as shown in Fig. 4(b) in the main text.
Examples of these measurements can be seen in Fig.~\ref{fig:RabiFit}, with green points indicating a measurement that was taken near the center of the slot and blue points marking data taken over the metal of the mw line.
Red and orange lines denote the exponentially decaying cosine fits performed on each dataset.
Datasets are offset for clarity.

\section{$T_2^*$ Calculation}
Optically detected magnetic resonance (ODMR) measurements were performed at a variety of mw power settings, and the lower-frequency [111] peak was fit with a gaussian function to determine linewidth.
\begin{figure}
    \centering
    \includegraphics[width=0.7\linewidth]{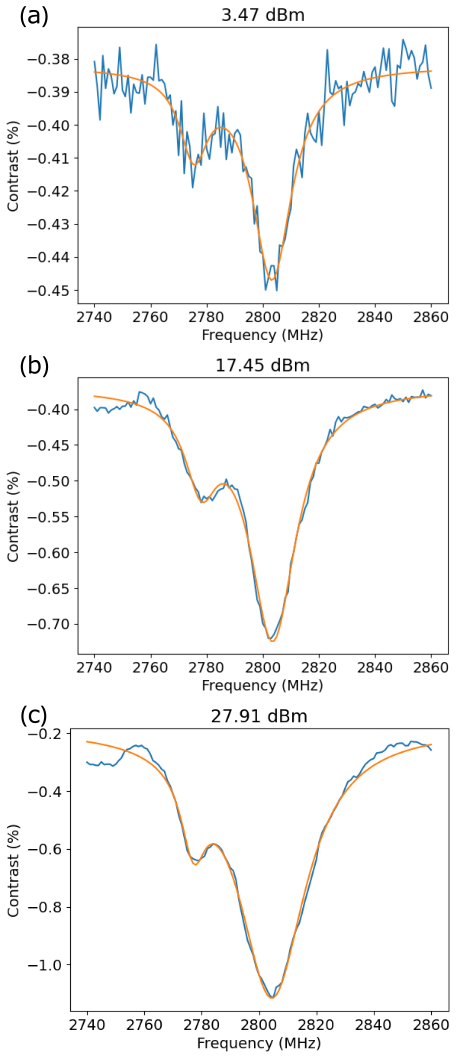}
    \caption{Example ODMR measurements at different mw powers, comparing experimental data in blue with gaussian fits in orange.  A two-peak gaussian fit was used to account for the smaller peak near~\SI{2780}{\mega \hertz}, but only the larger peak was used for calculations.}
    \label{fig:LWfit}
\end{figure}
Example measurements with fits are shown in Fig~\ref{fig:LWfit}, with experimental data in blue and gaussian fits in orange.
Though two peaks appear in these datasets, only the larger of the two is sensitive to magnetic field and therefore only the larger was used for calculations.
\begin{figure}
    \centering
    \includegraphics[width=0.9\linewidth]{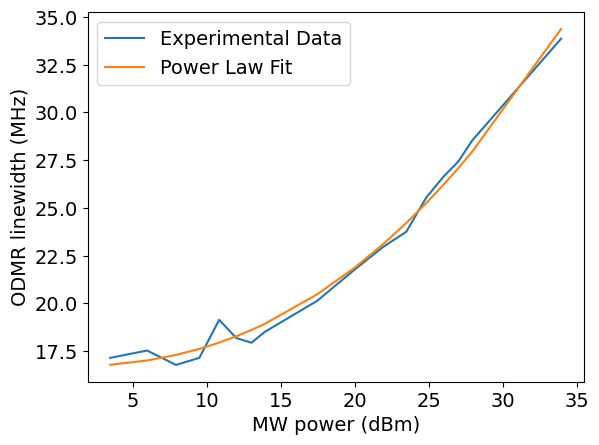}
    \caption{ODMR linewidth as a function of mw power in dBm.  Experimental data are shown in blue with a power law fit in orange.}
    \label{fig:LWvsMWgain}
\end{figure}
Plotting the linewidths obtained from the fits in Fig.~\ref{fig:LWfit} as a function of mw power produces the plot shown in Fig.~\ref{fig:LWvsMWgain}.
A power law function was fit to this data and found to have an intercept of~\SI{16.7(1)}{\mega \hertz}.
Using the equation $T_2^* = 1/(\pi \gamma)$, in which $\gamma$ is the ODMR linewidth at zero mw power, we obtained a value of~\SI{19.1(1)}{\nano \second} for $T_2^*$.

\section{DAC Assembly}
    This section describes in detail our procedure for assembling the diamond anvil cell with MW line used for the NV measurements presented in the main paper.
    The cell itself is an Almax Diacell SymmDAC 60 BeCu modified with a slot to accommodate the PCB that connects the MW line to an external MW source.
    Though the listed steps are specific to our experimental setup, many of the techniques described may be useful for other setups and DAC work in general.
        
    \subsection{PCB}
        The PCB depicted in Fig. 1(b) in the main text can be manufactured from simple single-sided~\SI{0.8}{\milli \meter}-thick FR4 board material using a PCB mill or laser cutter.
        Though the precise thickness is not important, the PCB must be thin enough that the top of the board remains below the diamond culet when both are mounted onto a backing plate.
        Otherwise, the gasket needed to hold the sample and pressure medium may rest on the PCB and be unable to contact the diamond.

        The ring-like end of the PCB is meant to sit on the backing plate around the NV-implanted diamond.
        The Pt MW line will be soldered to this end of the PCB in direct contact with the diamond, ensuring the NVs experience the strongest possible MW signal.
        An edge-mount SMA connector is soldered to the opposite end to connect the MW line to an external MW source.
        The type of solder does not matter as long as an electrical connection between the board and connector can be maintained under experimental conditions.
        
    \subsection{Mounting Diamonds}
        Each DAC requires two diamonds and two backing plates (seats).
        One of the diamonds should be implanted with NVs, and for the best performance, both should have the same culet diameter.
        Also ensure that the backing plates are non-magnetic, or they will interfere with NV measurements.
        Clean diamond tables, culets, and backing plate surfaces using cotton swabs or kimwipes soaked in ethanol or isopropanol.
        Fine-grit sandpaper can be used to remove leftover epoxy if necessary.

        \begin{figure}
            \centering
            \includegraphics[width=\linewidth]{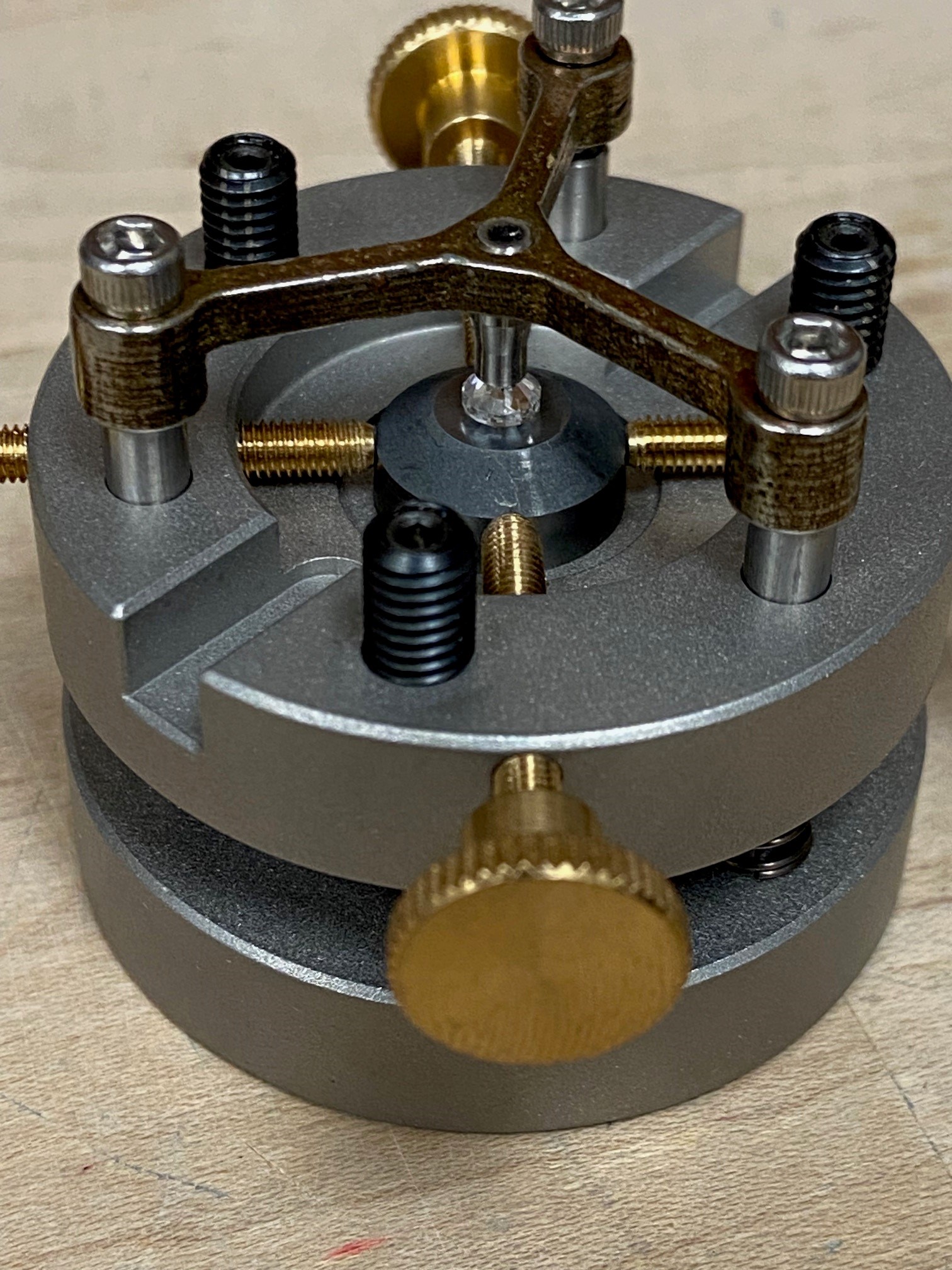}
            \caption{Picture of a mounting jig with a diamond and backing plate in place and ready for alignment.}
            \label{fig:jig}
        \end{figure}
        Once the diamonds and backing plates are clean, place one diamond and one backing plate together in each mounting jig, as shown in Fig.~\ref{fig:jig}.
        Also place the MW PCB around the NV-implanted diamond on its jig.
        The jig applies light pressure to the diamond and backing plate, ensuring that epoxy does not flow between them as it cures.
        Before applying epoxy, however, each diamond must be centered on its backing plate.
        Place one jig containing the diamond and backing plate upside-down under a microscope with an under-light.
        Look through the microscope to see the diamond culet.
        The culet should be illuminated by the under-light and appear as a small, bright circle inside the circular hole of the backing plate.
        While looking through the microscope, use the horizontal brass-colored screws on the sides of the jig to adjust the position of the backing plate relative to the diamond until the diamond culet is centered within the hole in the backing plate.
        When the two are aligned, tighten the brass screws to hold the backing plate in place.

        \begin{figure}
            \centering
            \includegraphics[width=\linewidth]{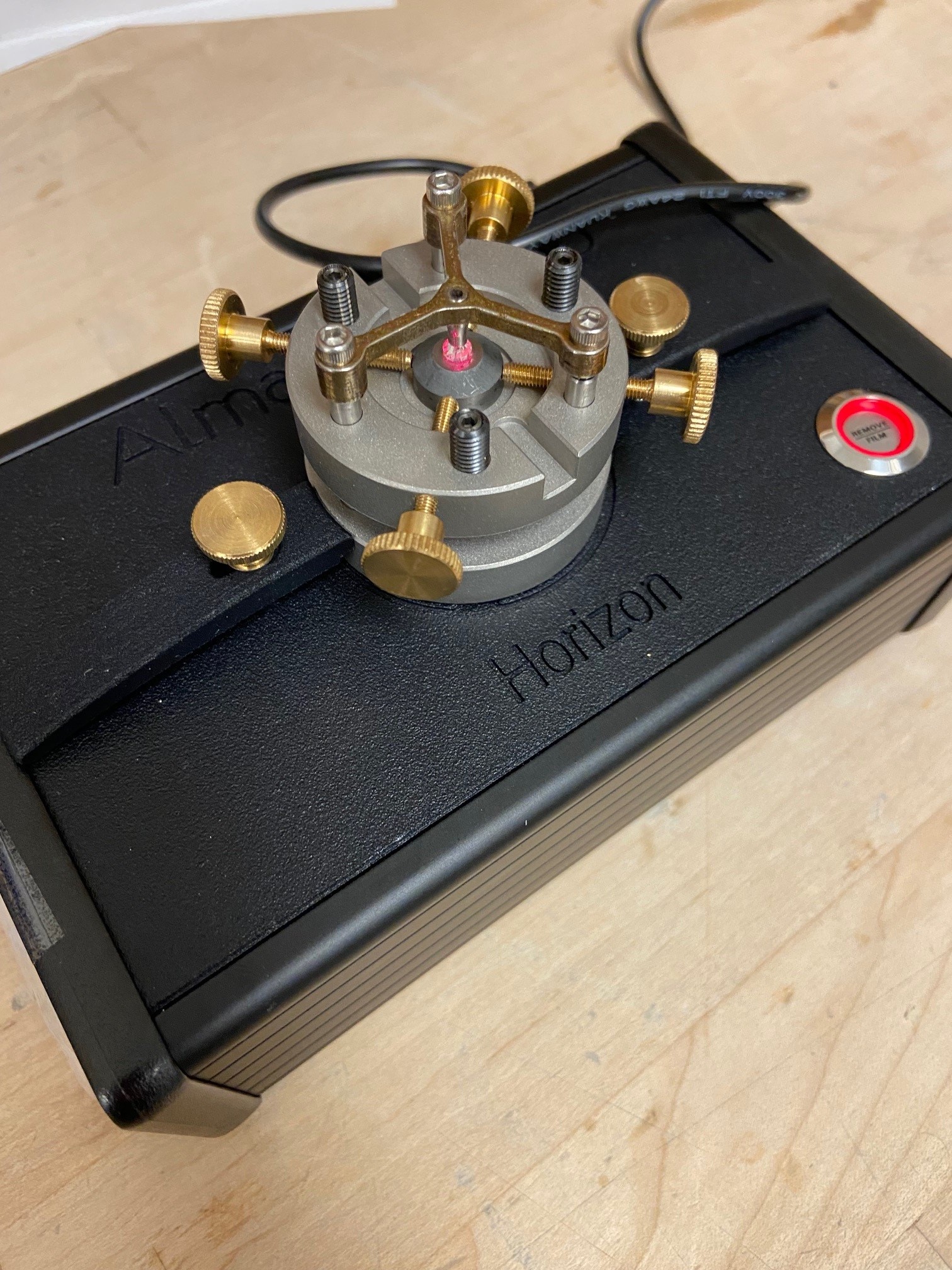}
            \caption{A mounting jig with diamond and backing plate placed on the Almax Horizon alignment tool.}
            \label{fig:AlignTool}
        \end{figure}
        To ensure the surface of the culet is horizontal, place this jig on the Almax Horizon laser alignment tool as depicted in Fig.~\ref{fig:AlignTool} and turn it on while aiming the laser at a blank point on the wall.
        If the diamond culet is not parallel with the base of the backing plate, two laser spots will be visible; otherwise, the spots will overlap and only one will appear.
        If two spots are seen, use tweezers to carefully rotate the diamond on the backing plate until the two spots are as close together as possible.
        When finished, turn the vertical black screws to firmly hold the backing plate and diamond together and remove the jig from the alignment tool.

        \begin{figure}
            \centering
            \includegraphics[width=\linewidth]{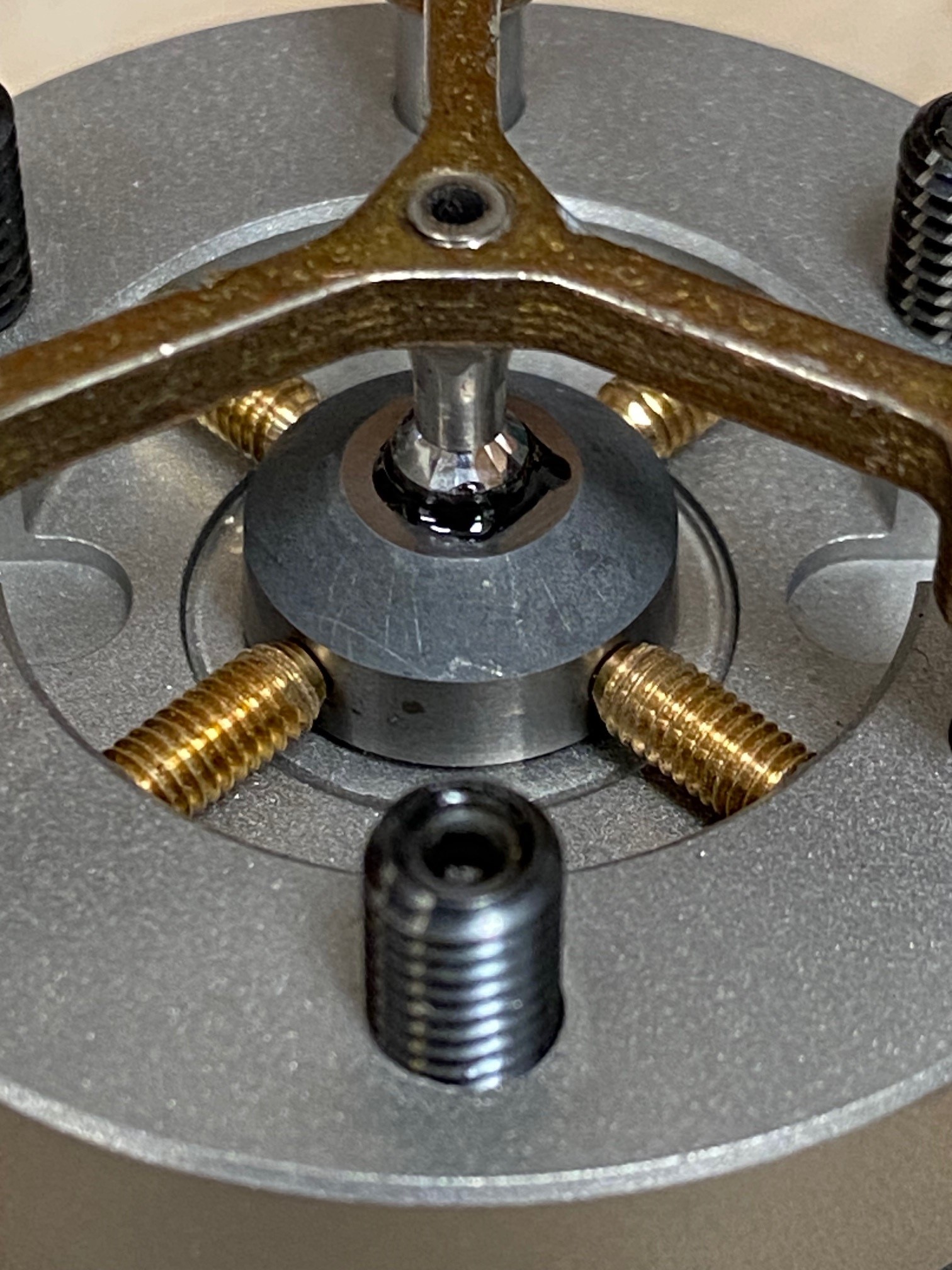}
            \caption{A ring of Stycast 2850FT epoxy around the base of a diamond on a backing plate in a mounting jig.}
            \label{fig:DiamondEpoxy}
        \end{figure}
\begin{figure}
            \centering
            \includegraphics[width=0.45\linewidth]{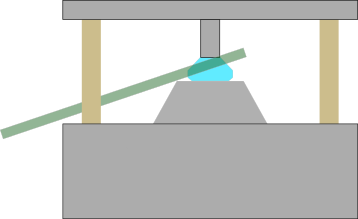}
            \includegraphics[width=0.45\linewidth]{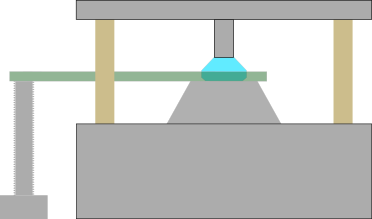}
            \caption{Diagrams of unsupported (left) and supported (right) PCBs in a mounting jig.}
            \label{fig:EpoxyPCB}
        \end{figure}

        Repeat the backing plate alignment with the remaining jig, then prepare a small amount of Stycast 2850FT epoxy by mixing it with a catalyst according to manufacturer directions.
        With the diamonds and backing plates still in their jigs, use a needle to apply a the epoxy in a small ring around the base of the diamond, as shown in Fig.~\ref{fig:DiamondEpoxy}.
        For the jig with the PCB, also apply a small amount of epoxy to the underside of the PCB.
        The PCB may not sit level on its backing plate; in this case, use a screw or similar small object to prop it up as demonstrated by Fig.~\ref{fig:EpoxyPCB}.
        Propping up the PCB will keep it fully in contact with the backing plate as the epoxy cures, ensuring a strong bond between the two and preventing the PCB from coming loose during measurement.

\begin{figure}
            \centering
            \includegraphics[width=0.8\linewidth]{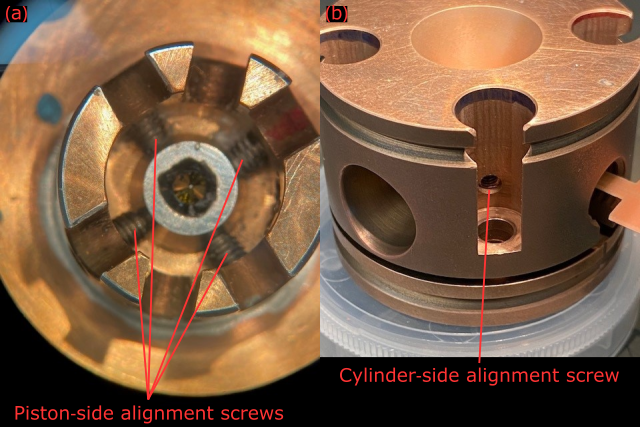}
            \caption{Picture of the two halves of a symmetric DAC showing the locations of the alignment screws used for holding and adjusting the positions of backing plates within the cell.}
            \label{fig:alignScrews}
        \end{figure}

        Cure the epoxy according to manufacturer directions, then remove the diamonds and backing plates from the jigs and place them in a diamond anvil cell, lightly tightening the alignment set screws to hold the backing plates in place.
        Figure~\ref{fig:alignScrews} shows where these screws are on each side of the cell.
        The backing plate with the NV-implanted diamond and PCB should do on the cylinder side of the cell, and the non-implanted diamond should be placed on the piston side.
            
    \subsection{Diamond-diamond Alignment}
\begin{figure}
            \centering
            \includegraphics[width=\linewidth]{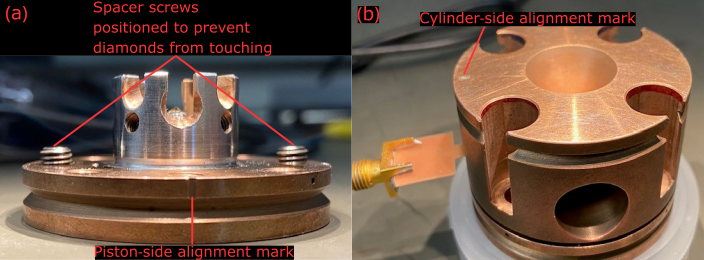}
            \caption{Picture of the two halves of a symmetric DAC showing the spacer screws and alignment marks.}
            \label{fig:spacer}
        \end{figure}

        The previous section included instructions on aligning each diamond with respect to its backing plate, but aligning the diamonds in the cell relative to each other is equally if not more important.
        Misalignment in this respect will make the gasket more likely to fail at lower pressures.
        Begin by turning the spacer screws on the piston side of the cell so that they emerge from the back of the piston as shown in Fig.~\ref{fig:spacer}(a).
        These screws control the spacing between the diamonds and will prevent them from touching one another, as diamond-on-diamond contact could cause them to crack or break.
        Therefore, these screws should emerge far enough that when the cell is closed, the diamonds are prevented from touching one another.
        Tighten the piston-side alignment screws very firmly, as these screws will not be adjusted further and will help to prevent the alignment from drifting in later steps when pressure is applied.
        Close the cell, making sure that the alignment marks depicted in Fig.~\ref{fig:spacer} on both halves are pointed in the same direction.
        Even after the diamonds are aligned, they will not be perfectly centered in the cell.
        Therefore, if the two halves of the cell are rotated relative to each other, the diamonds will no longer be aligned.
        \begin{figure}
            \centering
            \includegraphics[width=0.9\linewidth]{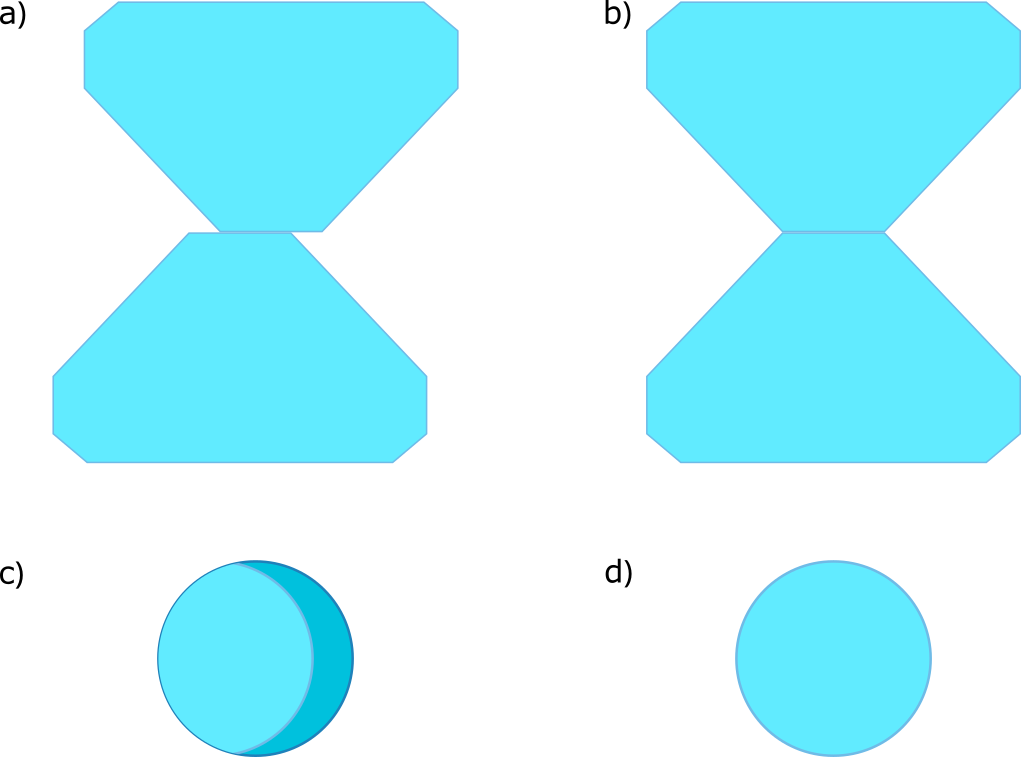}
            \caption{Simplified diagram of diamond alignment.  (a) Example of misaligned diamonds as viewed from the side.  (b) Side view of properly aligned diamonds. (c) Example of misaligned diamonds as viewed through a microscope. (d) Microscope view of properly aligned diamonds.}
            \label{fig:DiamAlign}
        \end{figure}
        
        Look through the windows on the sides of the cell to gain a rough idea of the diamond alignment and turn the cylinder-side alignment screws to align the diamonds as closely as possible by eye.
        Figure~\ref{fig:DiamAlign}(a) and (b) shows a sample diagram of of misaligned and aligned diamonds, respectively, as seen from the side.
        Then place the cell in a microscope with a light below the cell shining up through the diamonds.
        Look through the microscope at the diamond culets.
        If they are far apart, the viewer may not be able to see both culets at once.
        In this case, carefully adjust the spacer screws to bring the diamonds closer together.
        Alternate between checking the cell windows and watching the diamond culets under the microscope to ensure that the diamonds do not suddenly hit each other.
        The diamonds should be close enough that both are visible under the microscope without changing the focus.
        Additionally, keeping the diamonds closer can make alignment easier, especially if the microscope view is not perfectly vertical.
        The diamonds can be touched together if necessary, but do so very gently, be especially careful with the cell while the diamonds are in contact, and separate them as soon as alignment is complete.
        Also be mindful of the alignment marks, as the two halves of the cell may rotate.
        While looking at the culets under the microscope, turn the cylinder-side alignment screws to adjust the alignment.
        Misaligned diamonds seen through the microscope will appear similar to the diagram shown in Fig.~\ref{fig:DiamAlign}(c), while properly aligned diamonds will appear as in Fig.~\ref{fig:DiamAlign}(d).
        Once the diamonds appear closely aligned, rotate the cell \ang{90}.
        A change in angle may reveal issues that were not immediately apparent.
        Also check the alignment with one eye closed, and then the other.
        When the diamonds appear aligned from all angles and with each eye, tighten the alignment screws as much as possible without misaligning the diamonds again.
        Gasket preindentation sometimes causes the diamonds to shift position, and keeping the alignment screws tight may help to prevent this.

    \subsection{Gasket Preindentation}
\begin{figure}
            \centering
            \includegraphics[width=0.8\linewidth]{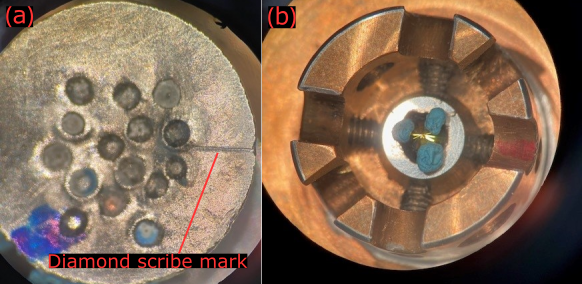}
            \caption{(a) Close-up of a used Re gasket showing the diamond scribe scratch used for orientation. (b) Blue sticky-tac placed around a diamond and ready to support the gasket.}
            \label{fig:Gasket}
        \end{figure}

        We use Re for a gasket material as it is stronger than steel and a single Re gasket can be reused multiple times.
        If using a new gasket, mark one edge with a diamond scribe; this mark, shown in Fig.~\ref{fig:Gasket}(a) will be used to maintain the same orientation of the gasket as it is removed from and placed back into the cell.
        On the piston side of the DAC, place three pieces of sticky tac around the diamond to support the metal gasket as demonstrated in Fig~\ref{fig:Gasket}(b).
        Clean the gasket, then place it on the diamond and sticky tac.
        It should be able to stand on its own.
        If not, adjust the sticky tac as necessary.
        If the gasket has been indented previously, ensure that the diamond is in contact with a fresh, non-indented part of the gasket.
        Mind the orientation of the diamond scribe mark on the gasket.
        Point the mark in a specific direction and/or indicate its direction in sharpie on the cell.
        \begin{figure}
            \centering
            \includegraphics[width=0.6\linewidth]{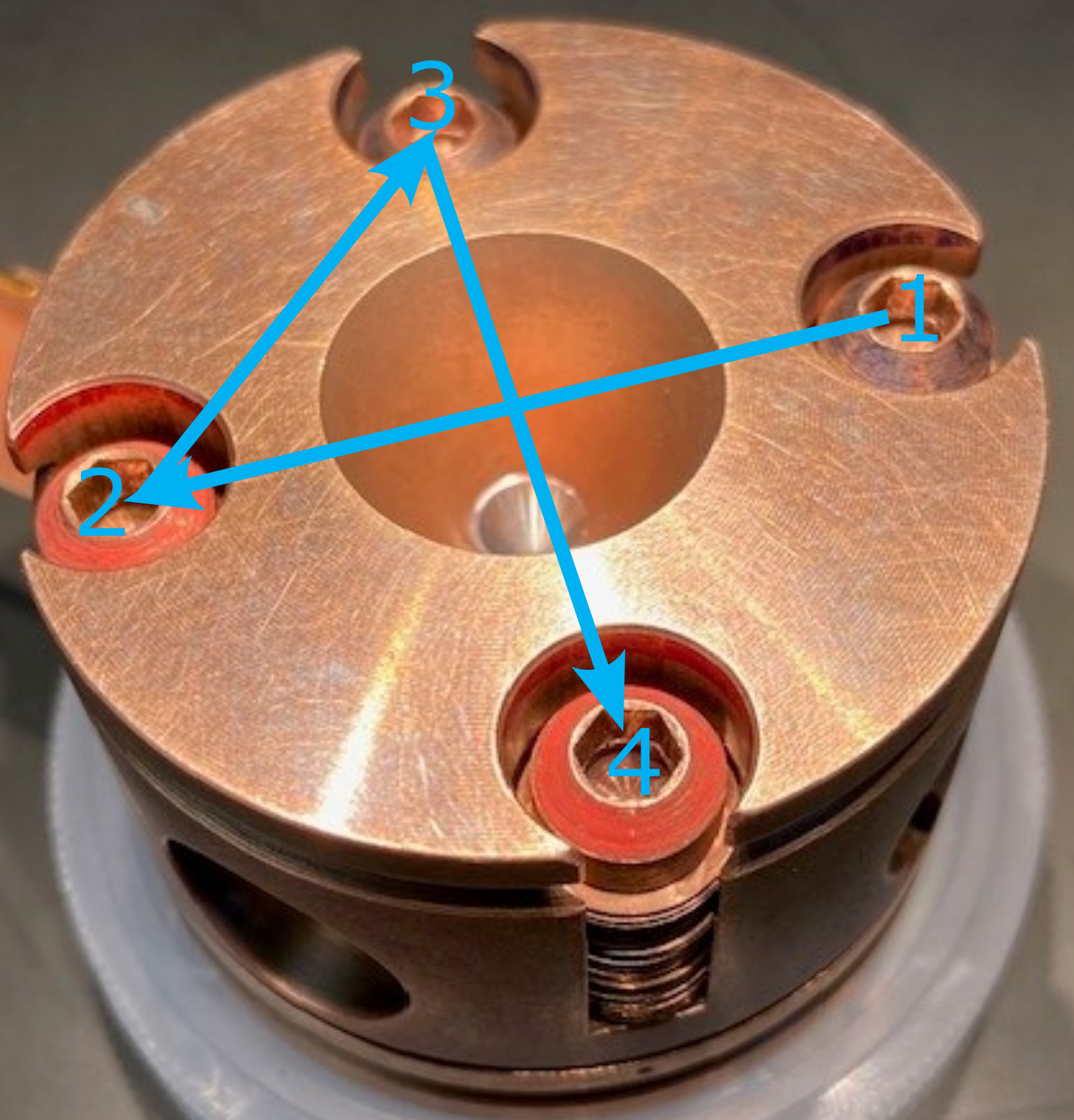}
            \caption{Closed symmetric DAC with bolts inserted.  Note the red and blue coloration on the bolt heads; these indicate left- and right-handed screws respectively and must be inserted into the correct hole.  Arrows and numbers represent an example 'x' pattern to use when tightening bolts to apply pressure.}
            \label{fig:CellWBolts}
        \end{figure}
        Close the cell and insert the bolts as depicted in Fig.~\ref{fig:CellWBolts}.
        These bolts will apply pressure when tightened, indenting the gasket.
        Using a torque wrench, turn each bolt until it reaches a maximum torque of about 0.5 in-lbs, moving from bolt to bolt in an 'x' pattern like the example given in Fig.~\ref{fig:CellWBolts}.
        Continuing to follow the same pattern, bring each bolt to a max torque of 1.0 in-lbs, then 1.5 in-lbs, and so on.
        When indenting and drilling a gasket, a useful rule of thumb is that the gasket hole diameter should be roughly 1/3 the diameter of the diamond culets, and the gasket thickness should be 1/3 of the hole diameter.
        For a Re gasket between~\SI{200}{\micro \meter} culets, continue until reaching a max torque of about 6.0 in-lbs.
        For a Re gasket between~\SI{300}{\micro \meter} culets, continue to a max torque of about 4.0 in-lbs.
        Then open the cell.
        If the gasket has been used previously, mark the current indentation in sharpie to keep track of it among other indentations.
        Remove the gasket and measure its thickness using a point micrometer.
        Though a gasket thickness of roughly 1/9 the culet diameter is desirable, keep in mind that the gasket insulation will add to its thickness.
        Therefore, at this point, aim for a gasket indentation thickness at least a few \si{\micro \meter} less than 1/9 the culet diameter.
        If the gasket is too thick, place it back into the cell as before and repeat the indentation process starting from 0.5 in-lbs but reaching a higher final torque.
        If the indentation is very thin, less than about 1/18 the culet diameter, begin again using a different indentation site or different gasket.
        Also check the alignment of the diamonds.
        Pressure from preindentation may move them out of alignment, in which case, they will need to be realigned and the gasket indentation process repeated.

    \subsection{Gasket Insulation}
\begin{figure*}
            \centering
            \includegraphics[width=\linewidth]{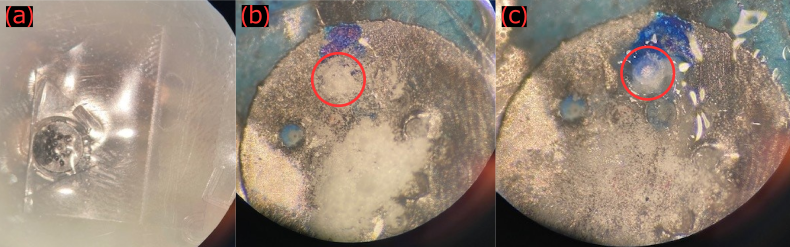}
            \caption{Steps of the gasket insulation process. (a) A piece of scotch tape folded over on itself with a gasket inside. (b) A pile of alumina powder on the gasket, with a small amount filling the indentation.  A red circle highlights the location of the indentation. (c) Compressed alumina powder in a gasket indentation surrounded by superglue.  A red circle highlights the location of the indentation.}
            \label{fig:Insulation}
        \end{figure*}

        Electrical contact between the conductive gasket and the Pt MW line will reduce MW transmission efficiency.
        Therefore the gasket must be insulated to maximize the signal-to-noise ratio of the ensuing NV measurements.
        Once the gasket has been preindented to an appropriate thickness, cover the entire gasket in a thin, clear tape such as scotch tape.
        We have had success using scotch tape as insulation, though we have not tested it at cryogenic temperatures.
        If measurements will be performed significantly below ambient temperature, other types such as kapton tape may be preferable.
        The tape may not stick to the gasket, so fold the tape over on itself as depicted in Fig.~\ref{fig:Insulation}(a).
        Using a razor blade or exacto knife, trim the excess tape from the edges of the gasket, leaving just enough for the tape to stick to itself, then cut out the area immediately around the indentation on both sides of the gasket.
        Place the gasket back in the cell as before, maintaining the same orientation that was used for preindentation.
        Scoop a small amount of alumina powder or stycast-alumina mix on to the gasket with a needle, toothpick, or fine-tipped tweezers, then push the powder into the preindented area until the indentation is filled, as shown in Fig.~\ref{fig:Insulation}(b).
        Close the cell and apply pressure using the same method as used to preindent the gasket, up to a max torque about 1 in-lbs below the max torque reached during preindentation.
        Open the cell and examine the now-insulated indentation.
        If the insulation does not reach all the way up the sides of the indentation, push more powder into the indentation and repeat the processes.
        Once the indentation is fully covered, apply a small amount of superglue in a ring around the edges of the compressed insulation powder with a needle.
        The superglue will hold the edges of the insulation powder in place, preventing pieces from breaking off as pressure is applied.
        The result of the completed insulation process is shown in Fig.~\ref{fig:Insulation}(c). 

    \subsection{Drilling Gasket \& Cutting MW Line}
        We use an LPKF Protolaser to cut the MW line and drill the gasket hole after insulation has been applied to the gasket.
        The Protolaser is not meant for single~\si{\micro \meter}-scale precision needed for gasket drilling, but this limitation can be overcome by calibrating the laser position using a blank test gasket or similarly-sized piece of metal immediately prior to drilling.
        Follow manufacturer directions to cut a hole into the center of the gasket indentation with a diameter approximately 1/3 the culet diameter.
        Keep in mind the laser beam diameter may produce a larger hole than desired.
        After drilling, examine the gasket indentation and hole, ensuring the hole is centered on both sides.
        If it is centered on one side but not the other, the diamond alignment has drifted and the diamonds will need to be realigned.
        If the gasket hole is off-center by less than 10\% of the culet diameter, the gasket should be usable but will be more likely to fail at lower pressures.
        The maximum pressure regularly achievable in diamond anvil cells can be approximated by the equation $P_{max} (GPa) = 12.5/d^2$, in which $P_{max}$ is the max pressure in~\si{\giga \pascal} and $d$ is the culet diameter of the diamonds in~\si{\milli \meter}~\cite{hamlin_superconductivity_2007}.
        If the gasket hole is not sufficiently centered, repeat the indentation, insulation, and drilling process.
        Otherwise, place the gasket back in the cell, maintaining its previous orientation, and close the cell.
        Insert the cell bolts and lightly tighten them to hold the gasket in place.
        Remove the sticky-tac supporting the gasket and replace it with a ring of superglue or 5-minute epoxy.
        The gasket should be attached to the piston side of the DAC.
        This adhesive will prevent the pressure medium from leaking out the bottom of the gasket and hold the gasket in place more robustly for easier loading.
        Open the cell again and examine the gasket hole.
        If any insulation has fallen into the hole, dig it out again using a fine-tipped needle.
        
        Also use the laser cutter to produce the slotted MW line shown in Fig. 1(b) from a piece of~\SI{4}{\micro \meter}-thick Pt foil.
        The thin Pt foil will ablate more easily than the thicker insulated gasket, so the same laser settings may remove a larger area of material, increasing the effective beam diameter.
        When the MW line is finished cutting, carefully solder it to the MW PCB as depicted in Fig. 1(c) in the main text.
        The angle of the MW line is not important, but the slot must be precisely placed over the center of the diamond culet.
        If the Pt MW line is springy and difficult to manipulate, it can be softened by annealing it with the flame of a lighter or in an oven.
        Once the MW line is soldered in place, do not close the cell until it is ready to measure, as the MW line is fragile and reopening the cell is likely to break it.
        
    \subsection{DAC Loading}
        The process for loading the cell with sample, Si oil pressure medium, and potentially a pressure marker such as ruby depends on the sample itself.
        If the sample is particularly light, pressure medium may be added first, as otherwise the flow of the pressure medium into the gasket hole may carry the sample away.
        If not, the sample may be added first, then the pressure medium, as manipulating the position of the sample may be easier without Si oil present.

        To place a solid piece of sample in the gasket hole, use a needle or eyelash tool.
        To make an eyelash tool, pluck out an eyelash and glue or epoxy it to the end of a toothpick or the handle of a cotton swab.
        Rub the side of your nose with the tip of your finger to collect some skin oil.  
        Gently touch the tip of the needle or eyelash tool to the tip of this finger.  
        This should make the tool sticky enough to pick up the sample but not so sticky that it cannot be placed in the gasket hole.  
        If the sample does not easily detach from the tool, gently scrape the sample against the side of the gasket hole.
        Once the sample is in the gasket hole, clean the tool with ethanol or isopropanol so that it does not accidentally pick up the sample again.  
        With the clean tool, adjust the sample position to place it roughly in the center of the hole.

        For a powder sample, clean any loose insulation powder from the gasket, then deposit a small amount of sample powder on the gasket.
        Use a needle to push a small amount of sample into the gasket hole, being careful not to fill it completely or push insulation powder into the hole.
        If the gasket hole is packed with sample and/or other solid material, there will not be enough room for the pressure medium, increasing pressure anisotropy within the cell.
 
        To add the Si oil pressure medium, dip a pair of fine-tipped tweezers into the Si oil, holding the tips close together to withdraw a single drop of oil, then touch them to the gasket to deposit the oil.
        If the oil does not flow onto the gasket, use a needle to draw the oil droplet away from the tweezers and onto the gasket.
        Even a single small drop of oil will likely be enough to cover the entire gasket including the sample space.
        Excess oil can be wicked away using a cotton swab or the tip of a Kimwipe if necessary.
        Also check for any bubbles in the gasket hole; if any exist, poke them with a needle to cause them to rise to the surface or pop.

        We do not use ruby for all measurements, as pressure can also be determined by ODMR frequency, and ruby fluorescence may increase background photon counts observed by the photodiode used for NV measurements, decreasing ODMR contrast.
        If using ruby for pressure determination, pick up a piece of ruby using a needle or eyelash tool with a very small amount of skin oil on it, the same way as described for a solid sample.
        Touch the surface of the Si oil with the tip of the tool with the ruby stuck on it.
        The ruby should fall into the Si oil.
        If not, gently scrape the ruby against the surface of the gasket.  
        Clean the tool, then push the ruby into the gasket hole.
        Adjust its position with an eyelash tool so that it is not obscured by sample.

        Once the pressure medium, sample, and optionally ruby are inside the gasket hole, double-check the position of the slotted MW line and adjust it if necessary.
        Then, keeping the gasket upright so that the pressure medium and sample do not flow out, close the cell and insert the bolts.
        Place the closed cell under a microscope to ensure the slot of the MW line has not moved from the center of the culet.
        If it has, carefully open the cell and adjust or replace the MW line as necessary, then close it again.
        With the cell closed and the MW line in position, apply a small amount of torque (0.5 in-lbs or less) to the bolts to ensure the pressure medium does not leak out.
        The cell is now ready for high-pressure NV measurements.
        When preparing to measure, remember that the NV-implanted diamond is on the cylinder side of the DAC.
        
\section{References}
\bibliographystyle{unsrt}
\bibliography{zotero_references,references}